\documentclass[aps,prl,twocolumn,showpacs,preprintnumbers,amsmath,amssymb,superscriptaddress]{revtex4-1}

\usepackage{graphicx}
\usepackage{dcolumn}
\usepackage{bm}
\newcommand{\nc}{\newcommand}
\nc{\be}{\begin{equation}}
\nc{\ee}{\end{equation}}
\nc{\bea}{\begin{eqnarray}}
\nc{\eea}{\end{eqnarray}}
\nc{\bean}{\begin{eqnarray*}}
\nc{\eean}{\end{eqnarray*}}
\nc{\mb}{\mbox}
\nc{\rnc}{\renewcommand}
\nc{\vk}{\mb{\bf k}}
\nc{\vp}{\mb{\bf p}}
\nc{\vn}{\mb{\bf n}}
\nc{\vq}{\mb{\bf q}}
\nc{\rr}{\mb{\bf r}}
\nc{\vz}{\hat {\mb{\bf z}}}
\nc{\vj}{\mb{\boldmath$j$}}
\nc{\vg}{\mb{\boldmath$g$}}
\nc{\x}{\mb{\boldmath$x$}}
\nc{\A}{\mb{\boldmath$A$}}
\nc{\va}{\mb{\boldmath$a$}}
\nc{\vs}{\mb{\boldmath$\sigma$}}
\nc{\vpi}{\mb{\boldmath$\pi$}}
\nc{\nab}{\nabla}
\nc{\X}{\sf x}

\begin{document}

\title{Fractionalization via $\mathbb{Z}_{2}$ Gauge Fields at a Cold Atom Quantum Hall Transition}

\author{Yafis Barlas}
\affiliation{National High Magnetic Field Laboratory and Department of Physics, Florida State
University, FL 32306, USA}
\author{Kun Yang}
\affiliation{National High Magnetic Field Laboratory and Department of Physics, Florida State
University, FL 32306, USA}

\begin{abstract}
We study a single species of fermionic atoms in an "effective" magnetic field at total filling factor $\nu_{f}=1$, interacting through a p-wave Feshbach resonance, and show that the system undergoes a quantum phase transition from a  $\nu_{f} =1 $ fermionic integer quantum Hall state to $\nu_{b} =1/4 $ bosonic fractional quantum Hall state as a function of detuning. The transition is in the $(2+1)$-D Ising universality class. We formulate a dual theory in terms of quasiparticles interacting with a $\mathbb{Z}_{2}$ gauge field, and show that charge fractionalization follows from this topological quantum phase transition. Experimental consequences and possible tests of our theoretical predictions are discussed.
\end{abstract}


\maketitle


\indent
Quantum number fractionalization in condensed matter systems is usually associated with topological order of the ground state. Although topological order inherent in fractionalization has been proposed in a number of systems including high-$T_{c}$ materials~\cite{Anderson,SenthilFisher} and spin liquids~\cite{moessnersondhi,spinliquids}, the only concrete example is in fractional quantum Hall (FQH) effect~\cite{Laughlin}. Ultra cold atoms provide an arena where topological order can be realized in quantum many body states by manipulating interactions. Interactions can be controlled by tuning the system through Feshbach resonances~\cite{Jin,Ketterle,pwave}. This allows one to investigate weak {\em and} strong paring regimes of fermionic superfluids with {\em both} s- and p-wave pairing interactions; the celebrated BCS-BEC crossover was clearly demonstrated for the former whereas the latter is expected to yield a rich phase diagram with a variety of phase transitions as a function temperature and interaction strength~\cite{pwave}. At the same time it has been suggested that FQH states can be realized in cold atoms by introducing "effective" gauge fields through rotation of the atomic trap~\cite{Fetter} or atom-light interaction~\cite{spielman}. Abrikosov vortex lattices have been observed in these systems both through trap rotation~\cite{Ketterle} and generation of synthetic gauge fields~\cite{spielman}. At sufficiently rapid rotational frequencies or subsequently low enough filling factor the single particle energies become highly degenerate and the system resembles an interacting electrons gas in a magnetic field; at this point the vortex lattice is expected to melt leading to various strongly correlated FQH states corresponding to different filling factors. In fact FQH like correlations have already been reported in an ensemble of rotating traps~\cite{GemelkeChu}. \newline
\indent
The possibility of fermion pairing in an external magnetic field~\cite{footnote1} allows one the unique opportunity to study Quantum Hall (QH) transitions driven by {\em attractive} interactions~\cite{haldane,qhtranskun}. In this paper we consider fermions interacting through a p-wave Feshbach resonance in an external magnetic field at total filling factor $\nu_{f} =1$. In the absence of interactions the fermions form an integer QH (IQH) state, which is stable against weak paring interactions. In the strong paring (or "BEC") limit two fermionic atoms combine to form a bosonic molecule with twice the effective charge $2e^{\star}$, and half the density of fermions; also the Landau level degeneracy is doubled due to the doubling of "charge", giving a bosonic filling factor $\nu_{b} =1/4$. These molecules thus can form a FQH state of the Laughlin type for repulsive interactions. Although both states have the same Hall conductance $\sigma_{xy} = (e^{\star})^2/h= (1/4)[(2e^{\star})^2/h]$, they exhibit different topological orders and must be separated by a phase boundary. We use Chern-Simons-Landau-Ginzburg (CSLG) theory~\cite{cslg} for QH states to show that this quantum phase transition (QPT) can be of second order in the $(2+1)$-D Ising universality class. To reveal the topological nature of the phases and phase transition, we perform a duality transformation to show that the low-energy theory of the system is a Chern-Simons/$\mathbb{Z}_{2}$ lattice gauge theory, coupled to a massive quasiparticle field. The transition occurs in the $\mathbb{Z}_{2}$ sector, which is also a confinement-deconfinement transition for the quasiparticles. As a result quasiparticle charge fractionalization and corresponding change of statistics angle accompany this transition, as well as changes of other topological properties like ground state degeneracy on high genus Riemann surfaces~\cite{WenandNiu}.
\newline
\indent
To study this QPT our point of departure is the CSLG theory, which relies on the off-diagonal long range order property of Laughlin type FQH states~\cite{allanandgirvin} by mapping them to charged superfluids coupled to a Chern-Simons (CS) gauge field. In the weak paring limit we can view the fermionic IQH state in the following way: attach one quantum of CS statistical flux to each fermionic atom thereby transforming it into a composite boson; at the mean field level the flux attached to the fermion cancels the external magnetic field, and the composite bosons see zero net flux. The fermionic IQH state then corresponds to the Bose condensed state of the composite bosons with charge-$e^*$ and can be regarded as an atomic superfluid (ASF) albeit with a CS interaction. In the strong pairing limit two charge-$e^{\star}$ bosons form a charge-2$e^{\star}$ boson bound state; each charge-2$e^{\star}$ boson carries two units of CS flux quanta and the bosonic FQH state corresponds to the charge-2$e^{\star}$ boson condensate and can be regarded as a molecular superfluid (MSF) with a CS interaction. As we show below, this effective CSLG field theory also properly describes the QPT between these two phases.  \newline
\indent
Taking $\hbar = c = e^{\star}=1$, the imaginary-time action of the theory can be written as
\bea
\label{fulllag}
& & \mathcal{S} = \int d \tau d^{2}x (\mathcal{L}_{0} + \mathcal{L}_{CS} + \mathcal{L}_{int}); \\
\nonumber
\mathcal{L}_{0} &=& \sum_{\sigma = 1,2} \bigg[\psi_{\sigma}^{\dagger} (\partial_{\tau} - \sigma a_{0})\psi_{\sigma} - \mu_{\sigma}|\psi_{\sigma}|^2  \\
\nonumber
&+& \frac{1}{2\sigma m}|(-i {\bf \nabla} - \sigma ({\bf A} + {\bf a}))\psi_{\sigma}|^2 + \cdots \bigg]; \\
\nonumber
\mathcal{L}_{int} &=& g(\psi^{\dagger}_{2}\psi_{1}\psi_{1}+ \psi^{\dagger}_{1} \psi^{\dagger}_{1}\psi_{2}); \quad
\mathcal{L}_{cs} = \frac{1}{4 \theta}  a_{\mu} \epsilon_{\mu \nu \lambda} \partial_{\nu} a_{\lambda}.
\eea
Here $\psi_{\sigma}$ are the bosonized atomic ($\sigma =1 $) and the molecular ($\sigma = 2$) fields, $\mu_{\sigma}$ is the respective chemical potential with $\mu_{2} = 2 \mu_{1} - \delta$ where $\delta$ is the detuning; positive detuning favors atoms while negative detuning favors molecules with the Feshbach resonance corresponding to $\delta =0$. $m$ is the mass of the atom and $g$ is the atom-molecular conversion matrix element, with the background density-density interactions implicit in "$\cdots$". The external magnetic field is incorporated in the vector potential ${\bf A}$, while ${\bf a}$ is the CS gauge field that attaches one (two) units of flux quanta to the atomic (molecular) fields respectively. The statistics angle $\theta = \pi$ encodes the flux attachment to the fermionic atoms transforming them into composite bosons.  The CS statistical gauge field (restoring fundamental units $\phi_{0} = hc/e$),
\be
{\bf \nabla} \times {\bf a} = \phi_{0} (|\psi_{1}|^2 +2|\psi_{2}|^2),
\ee
is chosen to cancel the external magnetic field ${\bf \nabla} \times (\langle {\bf a} \rangle + {\bf A}) =0$ on average; thus all bosons experience zero net flux. \newline
\indent
The action (\ref{fulllag}) defined above is invariant under the following local gauge transformations:
\be
\label{eqgt}
\psi_{1} \to \psi_{1} e^{i \theta(x,t)}; \
\psi_{2} \to \psi_{2} e^{2 i \theta(x,t)}; \
a_{\mu} \to a_{\mu} + \partial_{\mu} \theta(x,t).
\ee
The IQH state can be viewed as a charged Bose condensate of atoms {\em and} molecules: $\langle \psi_{1} \rangle \neq 0$ and $\langle \psi_{2} \rangle \neq 0$; thus the local $U(1)$ symmetry is "broken" (in an abuse of the term; see later for clarification) resulting in a mass to the gauge fluctuations $\Delta a_{\mu} = a_{\mu} + A_{\mu}$ through the Anderson-Higgs mechanism. The FQH state can be viewed as a charged Bose condensate of molecules $\langle \psi_{2} \rangle \neq 0$ while the atoms have not as yet condensed: $\langle \psi_{1} \rangle = 0$, again resulting in a mass for $\Delta a$.
Physically these reflect the fact the system is in an incompressible QH state throughout the phase diagram and charge fluctuations (which give rise to $\Delta a$) are suppressed.\newline
\indent
Despite the ordering of $\psi_2$, in the FQH state a discrete $\mathbb{Z}_{2}$ degree of freedom remains unspecified due to the $\pi$-periodicity of $\psi_{2}$ in (\ref{eqgt}). This is physically due to the fact that molecules are made of two atoms; thus specifying the phase of $\psi_2$ only determines the phase of $\psi_1$ modulo $\pi$. Hence the FQH state only breaks the $U(1)/\mathbb{Z}_{2}$ symmetry spontaneously, while the IQH state in which the phase of $\psi_1$ orders further breaks the remaining $\mathbb{Z}_{2}$ symmetry. Based on symmetry grounds it is easy to anticipate that an Ising-like phase transition must intervene between the $\nu_{b}=1/4$ FQH and $\nu_{f}=1$ IQH states in which this hidden $\mathbb{Z}_{2}$ symmetry is spontaneously broken. In the following we show that this is indeed the case and derive the effective theory for the phase transition. \newline
\indent
To work out the effective theory describing the transition we start in the $\nu_{b} = 1/4$ FQH region of the phase diagram where as we have already mentioned the molecules Bose condense $\langle \psi_{2} \rangle \neq 0$ (assumed real and positive without losing generality) thereby assigning a Higgs mass to the gauge fluctuations: $(\langle \psi_{2} \rangle^2/4m) |{\bf \Delta a}|^2$. This allows us to safely integrate out the massive CS gauge fluctuations $\Delta a_{\mu}$. To make the Ising degree of freedom transparent we perform a transformation on the resulting expression writing in terms of real scalar fields $\psi_{R}$ and $\psi_{I}$ with $ \psi_{1} = \psi_{R} + i \psi_{I}$,
\bea
\label{intefflag}
\nonumber
\mathcal{L}_{eff} &=&  2 i \psi_{R} \partial_{\tau} \psi_{I} + \frac{1}{2m} |\nabla \psi_{R}|^2 + \frac{1}{2m} |\nabla \psi_{I}|^{2}   \\
&-&(\mu_1 - 2h)|\psi_{R}|^{2}-(\mu_1 + 2h ) |\psi_{I}|^{2} + \cdots,
\eea
where $h = g\langle \psi_{2} \rangle$. Since $\mu_1 + 2h> \mu_1 -2h$, $\psi_{I}$ reaches criticality before $\psi_{R}$, the latter remains massive at the transition point and therefore can be safely integrated out of (\ref{intefflag}) to give us the effective theory for the transition,
\be
\mathcal{L}_{eff} =\frac{|\partial_{\tau} \psi_{I}|^2}{ 2h-\mu_1} + \frac{1}{2m} |\nabla  \psi_{I}|^2
- (\mu_1 + 2h) |\psi_{I}|^2 + \cdots,
\ee
which belongs in the $(2+1)$-D Ising universality class. This effective theory resembles the neutral atomic superfluid to molecular superfluid transition~\cite{rembertandleo} addressed by others in ultra cold atoms. 
The $\nu_{b} =1/4$ FQH to $\nu_{f} =1 $ IQH phase transition is driven by the condensation of $\psi_{I}$ which happens at $\delta_{c} \sim -4 h$; note that the presence of the molecular Bose condensate requires that $\mu_{1} \geq \delta/2$. \newline
\indent
While CSLG theory correctly describes the phase transition using the language of conventional Ginzburg-Landau theory (in terms of broken symmetry and order parameters etc), it is important to note that there is no {\it real} symmetry change between the $\nu_{b} =1/4 $ FQH state and the $\nu_{f} =1$ IQH state. This is because the order parameters of the CSLG theory, $\psi_1$ and $\psi_2$, are {\em not} gauge-invariant quantities. In fact
this is a phase transition between two {\em topological phases}, with different topological quantum numbers. For example, the ground state degeneracy on the torus is 1 for the  $\nu_{f} =1$ IQH state whereas  $ \nu_{b} =1/4$ FQH state has degeneracy $4$. To capture the topological nature of this QPT and correctly describe the change of topological properties of the system we must perform a duality transformation~\cite{fisherandlee} on (\ref{fulllag}).
\newline
\indent
To formulate the dual theory we work with a discrete version of (\ref{fulllag}) on a square lattice, the
lattice choice is arbitrary and we only require that the lattice version correctly reproduce (\ref{fulllag}) in the continuum limit. The lattice hamiltonian can be expressed in the rotor formalism in terms of noncommuting $[\hat{\theta}_{i,\sigma}, \hat{n}_{j,\sigma}] = i \delta_{ij}$ number $\hat{n}_{i,\sigma}$ and phase $\hat{\theta}_{i,\sigma}$ operators. The imaginary-time partition function is obtained in the usual way by discretizing imaginary time into $M-1$ steps of $\Delta \tau = \beta/M$. Inserting the closure relations associated to the phase and number fields at each time step we arrive at,
\begin{equation}
\label{pathint}
\mathcal{Z} =  \sum_{n_{1;i} n_{2;i}} \int \mathcal{D} \theta_{1}  \mathcal{D} \theta_{2} \mathcal{D} a_{\mu} \exp [S_{1} + S_{2} + S_{int} + S_{cs}],
\end{equation}
\begin{eqnarray}
\nonumber
S_{\sigma} &=& \sum_{i} \big[ n_{\sigma;i} (i \Delta_{\tau} \theta_{\sigma;i} - 2 \pi \sigma a_{0;i}) + \Delta \tau t_{\sigma} \sum_{\alpha} \cos (\Delta_{\alpha} \theta_{\sigma;i} \\
 &-& 2 \pi \sigma \delta {a}_{\alpha;i})  - \frac{\Delta \tau \lambda}{2} (n_{\sigma;i} - \bar{n}_{\sigma})^{2} + \cdots \big],
\end{eqnarray}
where $t_{\sigma}$ is the nearest neighbor hopping parameter for $\sigma =1(2)$ atoms (molecules), the integers $n_{\sigma;i}$ represent the atomic or molecular boson number and $ \bar{n}_{\sigma} = 1/2+(\mu_{\sigma})/\lambda$ and $\lambda >0 $ is the onsite repulsion. The lattice derivative is defined as $\Delta_{\mu} \theta_{\sigma;i} \equiv \theta_{\sigma;i+\mu} - \theta_{\sigma;i}$ with $\theta_{\sigma;i}$ a $2 \pi $ periodic angular variable and $ i =(i_{\tau},i_{x},i_{y})$ representing sites on the spacetime lattice, $\alpha = \hat{x},\hat{y}$ corresponds to spatial directions and $\mu = (\tau, \alpha)$ denotes the spacetime directions. The lattice version of the interaction and CS term is,
\begin{equation}
S_{int}= g \Delta \tau \sum_{i} \cos (2 \theta_{1;i} - \theta_{2;i}); \
S_{cs} = \frac{\pi}{2 \theta} \sum_{\Box} a_{\mu} \epsilon_{\mu \nu \lambda}  \Delta_{\nu} a_{\lambda}.
\end{equation}
To perform the duality transformation it is convenient to make the Villain approximation for the cosine terms in (\ref{pathint}), for example
\begin{equation}
\label{villain}
\exp[t \cos(\Delta_{\alpha} \theta_{i})] \cong \sum_{l_{\alpha;i}} \exp[t_{V}(\Delta_{\alpha} \theta_{i} - 2 \pi l_{\alpha;i})^2],
\end{equation}
with $t_{V}$ a renormalized constant. Since (\ref{villain}) preserves the periodicity of $\theta$ this change in functional form should not alter the physics. Hereafter we denote the modified partition function as $\mathcal{Z}_{V}$ and drop the unimportant normalization constants without comment. Following standard manipulations~\cite{fisherandlee} we introduce Hubbard-Stratonovich fields $  J^{1}_{\mu;i},J^{2}_{\mu;i},\eta_{i} $  which allows us to integrate out $\theta_{\sigma,i}$ on all sites and intermediate times to arrive at the current representation of the Villain approximation of (\ref{pathint}),
\bea
\label{currentrep}
\nonumber
\mathcal{Z}_{V} &=& \int \mathcal{D} a_{\mu} \sum_{\{L_{\mu;i}\}}  \exp \bigg( \sum_{i,\mu} \big[-\frac{(L^{1}_{\mu;i})^2}{2 \Delta \tau t_{1}} - \frac{(L^{2}_{\mu;i})^2}{2 \Delta \tau t_{2}} \\ \nonumber
&-&  \frac{(\eta_{i})^2}{2 \Delta \tau g} - 2 \pi i \Gamma_{\mu;i} \delta a_{\mu;i} \big]
+ \frac{\pi}{2 \theta} \sum_{\Box} a_{\mu} \epsilon_{\mu \nu \lambda}  \Delta_{\nu} a_{\lambda} \bigg)\\& &  \prod_{i} \delta(\Delta_{\mu} J^1_{\mu;i} - 2 \eta_{i}) \prod_{i} \delta(\Delta_{\mu} J^2_{\mu;i} -  \eta_{i}),
\eea
where $\{ L_{\mu;i} \} = \{ L^{1}_{\mu;i},L^{2}_{\mu;i},\eta_{i} \}$ with $L^{\sigma}_{\mu;i} \equiv J^{\sigma}_{\mu;i} - \bar{n}_{\sigma} \delta_{\mu \tau}$. $J^{\sigma}_{\mu;i} =(n_{\sigma;i},J^{\sigma}_{1;i},J^{\sigma}_{2;i})$ is the integer boson 3-current and we have chosen $\Delta \tau$ such that $t_{a} \Delta \tau = 1 /(\lambda_{a} \Delta \tau)$.  We transform to new link variables $\Gamma_{i \mu} = J^{1}_{i \mu} + 2 J^{2}_{i \mu}$ and $K_{i \mu} = J^{1}_{i \mu}$, modifying the constraints to $\Delta_{\mu} \Gamma_{\mu;i} =0$ and $\Delta_{\mu} K_{\mu;i} =2\eta_{i}$. The first constraint physically corresponds to the fact that the total atomic and molecular current $\Gamma_{i \mu} $ is conserved whereas the second constraint encodes the local atom-molecule interconversion events. The second constraint $\Delta_{\mu} K_{\mu;i} = 2 \eta_{i} $ can be used to sum over $\eta_{i}$. We can further sum over the link variables $K_{\mu;i}$. This summation is constrained as $K_{\mu;i}$ and $\Gamma_{\mu;i}$ have the same parity. To keep track of this parity, we divide $\Gamma_{\mu;i}$ into an even part and an "Ising" part which are {\it independently} conserved: $\Gamma_{\mu;i} =\Gamma^{e}_{\mu;i} + \Gamma^{I}_{\mu;i}$. Here $\Gamma^{e}_{\mu;i}$ is even while $\Gamma^{I}_{\mu;i}$ can only take values $-1,0,1$. This is always possible due to the total current conservation; links with odd $\Gamma_{\mu;i}$ must form loops, which can in turn be identified as loops of $\Gamma^{I}_{\mu;i}\ne 0$. Therefore in the summation $K_{\mu;i}$ has the same parity as that of $\Gamma^{I}_{\mu;i}$. After this summation, the partition function in terms of $\Gamma^{e}$ and $\Gamma^{I}_{\mu;i}$ takes the form,
\bea
\nonumber
\label{interme}
\mathcal{Z}_{V} &=& \int \mathcal{D} a_{\mu} \sum'_{\Gamma^{e}_{\mu;i},\Gamma^{I}_{\mu;i}}  \exp \bigg( \sum_{i,\mu} \big[-\frac{(\Gamma^{I}_{\mu;i})^2}{2 \Delta \tau t'_{1}} - \frac{(\Gamma^{e}_{\mu;i})^2}{8 \Delta \tau t'_{2}} \\  &-& 2 \pi i \Gamma_{\mu;i} \delta a_{\mu;i} \big]
+ \frac{\pi}{2 \theta} \sum_{\Box} a_{\mu} \epsilon_{\mu \nu \lambda}  \Delta_{\nu} a_{\lambda} + \cdots \bigg),
\eea
with the constraints $\Delta_{\mu} \Gamma^e_{\mu;i}=\Delta_{\mu} \Gamma^I_{\mu;i}=0$. They can be solved as usual by defining integer link variables $g_{\mu;a}$ and $h_{\mu;a}$ on a dual lattice,
\be
\Gamma^{e}_{i \mu} = \epsilon_{\mu \nu \lambda} \Delta_{\nu} g_{\lambda;a}, \qquad \Gamma^{I}_{i \mu} = \epsilon_{\mu \nu \lambda} \Delta_{\nu} h_{\lambda;a},
\ee
where $a$ denotes the sites of the dual cubic lattice and $g_{\mu;a}$ is an {\it even}-valued integer field and we define $h_{\mu;a}$ as an integer valued field. To proceed we remove the {\it even}-integer constraints on $g_{\mu;a}$ by performing a Poisson resummation through half-integer link variables $m_{\mu;a}$ and exploit the gauge invariance $\pi g_{\mu ;a} \to  \pi g_{\mu;a} - \Delta_{\mu} \varphi_{a}$ to introduce $\varphi_{a}$ which we later identify as the phase of the molecular vortex field. This choice is slightly different from convention~\cite{fisherandlee}, and reflects the evenness of $g_{\mu;a}$.
Adding a vortex fugacity term $1/y \sum_{\mu;a} |m_{\mu;a}|^2$ we can integrate out $m_{\mu;a}$ at the expense of another Possion resummation even-integer dual link variable $n_{\mu;a}$ to obtain
\bea
\nonumber
\mathcal{Z}_{V} &=& \int \mathcal{D} \varphi_{a} \mathcal{D} a_{\mu} \mathcal{D} g_{\mu} \sum_{n_{\mu;a},h_{\mu;a}}  \exp \bigg( \sum_{\Box} \big[ \frac{\pi}{2 \theta}  a_{\mu} \epsilon_{\mu \nu \lambda}  \Delta_{\nu} a_{\lambda}  \\ \nonumber
&+& 2 \pi i \epsilon_{\mu \nu \lambda} \Delta_{\nu} (g_{\lambda;a} + h_{\lambda;a}) \delta a_{\mu;i}-\beta \cos(\pi \epsilon_{\mu \nu \lambda} \Delta_{\nu} h_{\lambda;a}) \big]\\
&+& y \sum_{\mu}  (2\Delta_{\mu} \varphi_{a} - 2\pi g_{\mu;a} + 2 \pi n_{\mu;a} )^2 + \cdots   \bigg)
\end{eqnarray}
where  $\beta =1/(4 \Delta \tau t'_{1}) $, from here onwards the Maxwell and other higher order terms will be implicit in "$\cdots$". Redefining the $U(1)$ gauge field $ b_{\mu;a} = g_{\mu;a } + h_{\mu;a}$ in the resulting expression and integrating out the CS gauge field which is coupled to $b_{\mu;a}$ (whose curl is the {\em total} conserved current), one arrives at the dual version of (\ref{pathint}):
\begin{eqnarray}
\label{finaldual}
\nonumber
\mathcal{Z}_{V} &=& \int \mathcal{D} \varphi_{a} \mathcal{D} b_{\mu} \sum_{\sigma_{\mu;a}}  \exp \bigg( y \sum_{\mu} \sigma_{\mu;a} \cos( \Delta_{\mu} \varphi_{a} - \pi b_{\mu;a})\\
&-&\sum_{\Box} \big[\beta \prod_{\Box} \sigma_{\mu;a} +  4  \theta \pi  b_{\mu} \epsilon_{\mu \nu \lambda}  \Delta_{\nu} b_{\lambda} \big] + \cdots   \bigg),
\end{eqnarray}
where we have introduced the Ising gauge field $\sigma_{\mu;a} = \cos (\pi h_{\mu;a})$.  In (\ref{finaldual}) the angular variable $\varphi_{a}$ corresponds to the phase of the molecular vortex field. Due to the presence of the (dual) CS term such a vortex carries charge $e^{\star}/2=2e^*/4$ with $\pi/4$ statistics. It corresponds to the quasiparticle excitation of the $\nu_{b} =1/4$ bosonic FQH state. \newline
\indent
In $d+1 \geq 3 $ $\mathbb{Z}_{2}$ gauge theory exhibits a confinement-deconfinement phase transition as a function of $\beta$~\cite{z2theory}. The difference between the confining and deconfining phases relies on the notion of topological order which has important implications on the nature of excitation of the system. Pertinent to our case is the charge fractionalization that occurs in the deconfined phase: $e^{\star}/2 $ quasiparticles (molecular vortices) are free to propagate, this corresponds to the molecular FQH phase.
In the confined phase the energy grows linearly with the separation of two $e^{\star}/2 $ quasiparticles: they thus bind together and form charge-$e^{\star}$ with $4\times \pi/4$ or {\em fermionic} statistics, these are nothing but the original fermionic atoms. Thus the confined phase corresponds to the atomic IQH phase. A more physical way understand this transition is in terms of  condensation of visons, which are excitation of the $\mathbb{Z}_{2} $ flux through a plaquette ($\prod_{\Box} \sigma =-1$) ~\cite{SenthilFisher}. In the confined phase visons are condensed and particles carrying $\mathbb{Z}_{2}$ charge cannot propagate, whereas in the deconfined phase where the visons have not as yet condensed these topological excitations cost finite energy, and the theory is in the deconfined phase. Furthermore (\ref{finaldual}) correctly captures the topological degeneracy of the phases on high genus surfaces. Using torus as an example, this degeneracy is 4 in the deconfined phase as in the $\nu_{f} =1/4$ FQH state, and 1 in the confined phase as in the $\nu_{f} =1$ IQH state\cite{spinliquids,degeneracynote}. Thus the dual theory (\ref{finaldual}) captures {\em all} the topological properties of the phases, and properly describes the phase transition from $\nu_{b} =1$ IQH state to $\nu_{f} =1/4$ FQH state. \newline
\indent
For $d=2$ this confinement-deconfinement phase transition is in the $3$-D Ising universality class~\cite{z2theory}. The critical exponent of the $3$-D Ising model are well known and standard scaling arguments~\cite{sachdev} predict that a gap vanishes as one approaches the critical point: $\Delta \sim |\delta - \delta_{c}|^{z \nu}$, where $\nu\approx 0.63$ is 3D Ising correlation length exponent and the dynamical exponent $z=1$ due to (emergent) Lorentz invariance. Physically this gap corresponds to the long-wavelength magneto-phonon gap in the $\nu_{b} =1/4$ FQH state which can be detected by probing the excitation spectrum through simulated Bragg spectroscopy~\cite{braggspec}. \newline
\indent
The stability and observability of the FQH phase and the critical point depend on the repulsive nature of the background molecule-molecule interaction assumed here, and in particular, the stability of the atoms and molecules themselves. While in 3D their decay rate is comparable to the interaction scale near the p-wave resonance, the situation becomes much more favorable in quasi-2D systems~\cite{cooper}. The presence of an external "magnetic field" may further help by preventing multiple particles from coming close together. We also note that the transition occurs close to, but {\em not} precisely at the resonance; the distance between criticality and resonance depends on density and other parameters. This allows for additional room in searching for a parameter window where the critical point is stable.
\newline

This work was supported in part by NSF grant DMR-1004545 (KY)
and the State of Florida (YB).


\begin{thebibliography}{99}

\bibitem{Anderson}
P.W.Anderson, Science {\bf 235}, 1196 (1987).

\bibitem{SenthilFisher}
T.Senthil and M.P.A.Fisher, Phys. Rev. B {\bf 61}, 9690 (2000).

\bibitem{moessnersondhi} R. Moessner and S. L. Sondhi, Phys. Rev. Lett. {\bf 86}, 1881 (2001).

\bibitem{spinliquids}
See {\em e.g.}, X.G.Wen, Quantum Field Theory of Many-Body Systems, (Oxford University Press, Oxford 2004).

\bibitem{Laughlin}
R.B.Laughlin, Phys. Rev. Lett. {\bf 50}, 1395 (1983).

\bibitem{Jin}
C.A.Regal, M.Greiner and D.S.Jin, Phys. Rev. Lett. {\bf 92}, 040403 (2004).

\bibitem{Ketterle}
 M.W.Zwierlein, J.R.Abo-Shaeer, A.Schirotzek, C.H.Schunck and W.Ketterle, Nature (London) {\bf 435}, 1047 (2005).

\bibitem{pwave}
V.Gurarie, L.Radzihovsky and A.V.Andreev, Phys. Rev. Lett. {\bf 94}, 230403 (2005).

\bibitem{Fetter}
A.L.Fetter, Rev. Mod. Phys. {\bf 81}, 647 (2009).

\bibitem{spielman}
Y.-J.Lin, R.L.Compton, K.Jimenez-Garcia, J.V.Porto and I.B.Spielman, Nature {\bf 462}, 628 (2009).

\bibitem{GemelkeChu}
N.Gemelke, E.Sarajlic and S.Chu, cond-mat/1007.2677.

\bibitem{footnote1}
From now on we refer to the "effective" magnetic field whether generated through rotation or atom-light interaction as external.

\bibitem{haldane}
F. D. M. Haldane and E. H. Rezayi, KITP conference on Quantum Gases (2004).

\bibitem{qhtranskun}
K.Yang and H.Zhai, Phys. Rev. Lett. {\bf 100}, 030404 (2008).

\bibitem{cslg}
S.C.Zhang, H.Hanson and S.Kivelson, Phys. Rev. Lett, {\bf 62}, 82 (1989); 62, 980 (1989); S.C.Zhang, Int. Jour. Mod. Phys. B {\bf 6}, 25 (1992).

\bibitem{WenandNiu}
X.G.Wen and Q.Niu, Phys. Rev. B {\bf 41}, 9377 (1990).

\bibitem{allanandgirvin}
S.Girvin and A.H.MacDonald, Phys. Rev. Lett. {\bf 58}, 1252 (1987); N. Read, Phys. Rev. Lett. {\bf 62} 86 (1089).


\bibitem{rembertandleo}
L.Radzihovsky, J.Park and P.B. Weichman, Phys. Rev. Lett. {\bf 92}, 160402 (2004); Annals of Phys. {\bf 323}, 2376 (2008); M.W.J.Romans, R.A.Duine, S.Sachdev and H.T.C.Stoof, Phys. Rev. Lett. {\bf 93}, 020405 (2004).


\bibitem{fisherandlee}
M.P.A.Fisher and D.H.Lee, Phys. Rev. B {\bf 39}, 2756 (1989).



\bibitem{z2theory}
J.B.Kogut, Rev. Mod. Phys. {\bf 51}, 659 (1979); E.Fradkin and S.H.Shenker, Phys. Rev. D {\bf 19}, 3682 (1979).

\bibitem{degeneracynote} The U(1) sector with the CS term can also contribute to this degeneracy. Here it is 1 as the system is in the $\nu=1$ QH state in unit of the fundamental (atomic) charge.

\bibitem{sachdev}
S.Sachdev, Quantum Phase Transitions (Cambridge University Press, Cambridge 1999).


\bibitem{braggspec}
J. Stenger {\it et. al.}, Phys. Rev. Lett. {\bf 82}, 4569 (1999).

\bibitem{cooper}
J. Levinsen, N.R.Cooper and V.Gurarie, Phys Rev. B {\bf 78}, 063616 (2008).
\end{thebibliography}
\end{document}